\documentclass[jkps,twocolumn,amsart]{revtex4}
\usepackage[pdftex]{graphicx}
\usepackage{amssymb}
\usepackage{amsmath}
\usepackage{bm}

\newcommand{\etal}{{\it et al.}}
\newcommand{\EF}{$E_{\rm F}$}
\newcommand{\TC}{$T_{\rm C}$}
\newcommand{\ELB}{Eu$_{1-x}$La$_x$B$_6$}
\newcommand{\ELBE}{Eu$_{0.8}$La$_{0.2}$B$_6$}

\newcommand{\Gn}{$\overline{\Gamma}^{(n)}$}
\newcommand{\Xone}{$\overline{X}^{(1)}$}

\newcommand{\Xtwo}{$\overline{X}^{(2)}$}
\newcommand{\Xn}{$\overline{X}^{(n)}$}
\newcommand{\Mn}{$\overline{M}^{(n)}$}

\begin{document}
\setcounter{page}{0}
\title{Semimetallic Nature and Magnetic Polarons in EuB$_6$ Studied by Angle-Resolved Photoemission Spectroscopy}

\author{Chul-Hee \surname{Min}}
\affiliation{Institut für Experimentelle und Angewandte Physik, Christian-Albrechts-Universität zu Kiel, D-24098 Kiel, Germany}

\author{Boyoun \surname{Kang}}
\author{Beong Ki \surname{Cho}}
\affiliation{School of Material Science and Engineering, Gwangju Institute of Science and Technology, Gwangju 500-712, South Korea}

\author{En-Jin \surname{Cho}}
\affiliation{%
Department of Physics, Chonnam National University, Kwangju 500-757, South Korea}

\author{Byeong-Gyu \surname{Park}}
\affiliation{%
PLS-II Beamline Division, Pohang Accelerator Laboratory, Pohang 37673, South Korea}

\author{Hyeong-Do \surname{Kim}}
\email{hdkim6612@postech.ac.kr}
\affiliation{%
XFEL Beamline Division, Pohang Accelerator Laboratory, Pohang 37673, South Korea}

\date[]{Received \today}

\begin{abstract}
Temperature-dependent angle-resolved photoemission spectroscopy (ARPES) was carried out on single-crystalline EuB$_6$ samples. 
By measuring ARPES spectra in an extended Brillouin zone, a B 2$p$ hole pocket centered at the $X$ point is clearly observed, thus proving the semimetallic nature of EuB$_6$.
Below the Curie temperature \TC, ARPES spectra show two B 2$p$ bands of which separation is due to an exchange interaction between local Eu 4$f$ and itinerant B 2$p$ electrons.
The exchange splitting becomes smaller as the temperature increases and disappears well above \TC.
Additionally, a diffuse structure near the Fermi level survives just above \TC. 
Such behavior is well described by Monte Carlo simulations of a Kondo lattice model, thus supporting the formation of magnetic polarons in EuB$_6$, which accounts for the resistivity upturn near above \TC\ when lowering the temperature.
\end{abstract}

\keywords{EuB$_6$, semimetal, magnetic polaron, angle-resolved photoemission spectroscopy}

\maketitle

A low-carrier-density system EuB$_6$ has long attracted much attention because it shows an anomalous resistivity behavior concomitant with its ferromagnetic (FM) phase transition near \TC\ $\sim 15.5$~K \cite{Guy}. 
In addition, another phase transition related to electronic excitations has been revealed in the FM phase by specific-heat and optical reflectivity measurements \cite{Degiorgi}.
It has been proposed that the low carrier density is essential to stabilize the FM order in a Kondo lattice system \cite{Sigrist}.
According to the LDA or LDA+$U$ band calculations \cite{Massidda,Kunes,Shim}, the Eu 5$d$ and the B 2$p$ bands overlap each other and form small electron and hole pockets, respectively, centered at the $X$ point in the Brillouin zone (BZ), to produce low-density carriers. 
This picture is experimentally supported by Shubnikov-de Haas and de Haas-van Alphen (dHvA) measurements \cite{Goodrich,Aronson}, optical conductivity \cite{Kim05,Kim08}, Andreev reflection spectroscopy \cite{Zhang08}, and resonant inelastic x-ray scattering \cite{Kim13}.
On the other hand, $GW$ calculations for another divalent hexaboride CaB$_6$ show an energy gap at the $X$ point \cite{Tromp}, which is confirmed by angle-resolved photoemission spectroscopy (ARPES) \cite{Denlinger02,Souma}, while it is predicted to be semimetallic by LDA calculations \cite{Hasegawa,Tromp}.
ARPES spectra of EuB$_6$ obtained by Denlinger \etal\ also show a bulk band gap at the $X$ point \cite{Denlinger02}.
Recent, ARPES measurements of CeB$_6$ also show a similar overlap of the Ce 5$d$ and the B 2$p$ bands at the $X$ point, which is well described by theoretical bulk band structure \cite{Neupane}. 
Recent ARPES studies of EuB$_6$ focused on its topological properties show a negligible gap in the paramagnetic (PM) phase and clear semimetallic feature in the FM phase \cite{Gao21,Liu21}.
Tunneling measurements also seem to provide evidence for a bulk band gap of 43~meV \cite{Amsler}.
Thus, there are some controversial results on the origin of the low-density carriers in EuB$_6$.

Another interesting issue on EuB$_6$ is that when decreasing temperature, the resistivity shows an upturn at about 30~K and rapidly decreases down to about 10~K  \cite{Guy,Sullow98}, which seems to be related with two consecutive transitions in the FM phase \cite{Degiorgi}.
All the results from Raman spectra \cite{Nyhus,Snow}, neutron scattering \cite{Henggeler}, field-dependent resistivity and magnetization measurements \cite{Sullow00}, muon-spin rotation measurements \cite{Brooks}, electron-spin resonance measurements \cite{Urbano}, nonlinear Hall effects \cite{Zhang09}, and fluctuation spectroscopy and weak nonlinear transport measurements \cite{Das12} suggest that the phase transition at \TC\ is due to the overlap of bound magnetic polarons (MPs), thus inducing a percolative transition in the resistivity, and the lower-temperature one a bulk FM order.
Theoretical investigations based on a Kondo lattice model also suggest the formation of bound MPs \cite{Yu05}, and can reproduce the resistivity anomaly and the consecutive transitions, though the lower-temperature transition is percolative \cite{Yu06}.

In this work, we present APRES results of \ELB\ in an extended BZ to observe the B 2$p$ band crossing the Fermi level (\EF). 
Our data confirm the semimetallic nature of EuB$_6$, thus resolving the issue of the origin of the low-density carriers.
Furthermore, temperature-dependent ARPES spectra show two clear exchange-split bands in the FM phase and an exchange-driven diffuse structure near \EF\ even in the paramagnetic (PM) phase.
Such a structure is explained using a two-dimensional (2D) Kondo lattice model, which shows MPs near above \TC.

Single-crystalline samples of \ELB\ ($x= 0.0$ and 0.2) were grown by an Al flux method \cite{Rhyee06} and mounted on sample holders for ARPES on which their orientations were determined by their rectangular shapes or Laue pictures.
ARPES measurements were carried out at the 4A1 $\mu$-ARPES Beamline of PLS-II at Pohang Accelerator Laboratory equipped with a Scienta-SES2002 electron analyzer.
All samples were cleaved {\it in situ} to show a flat (001) surface by a top-post method in a vacuum better than $7 \times 10^{-11}$~Torr.
Photon energies from 100 to 150~eV were used to probe a three-dimensional (3D) electronic structure,  and no noticeable change was observed during the measurements.
The total energy and momentum resolutions were set to be about 100~meV and 0.01~\AA$^{-1}$, respectively, and \EF\ is determined from the Fermi edge of a Au film electrically connected to the samples.
Sample temperature was lowered by an open-cycle liquid-He cryostat and controlled by a resistive heater within 0.1~K.

Since there is a complicated issue related to the surface electronic structure of EuB$_6$ in ARPES spectra \cite{Denlinger02}, we first investigated the electronic structure of electron-doped \ELBE.
Since the number of electron carriers is big, its electronic structure may not be so much affected by surface defects as in a low-carrier-density system.
According to the previous ARPES studies \cite{Denlinger02}, the inner potential of EuB$_6$ was determined to be about 11.2~eV, which renders a value of $h\nu = 130$~eV to reach the $\Gamma$ point in the region of $h\nu = 100 \sim 150$~eV.
Figure~1(a) shows the Fermi-surface (FS) map of \ELBE\ at $h\nu = 130$~eV, which shows a large elliptical electron pocket centered at the $\overline{X}$ point projected on to the surface BZ.
To see how close $h\nu = 130$~eV accesses the $X$ point, we obtained FS maps by varying the photon energy from 115 to 146~eV, the results of which are shown in Fig.~1(b).
FS maps denoted by $\alpha$, $\beta$, and $\gamma$ are the cross-sectional views of the 3D ellipsoidal FS as shown in the upper left drawing.
Since $\alpha$ and $\gamma$ are for the same symmetry plane $\Gamma XM$, they should be the same but the FS shape of $\alpha$ looks a little distorted when compared to that of $\gamma$ (see blue dashed guidelines for the FSs).
This may originate from a 3D artifact and strong matrix-element effects in ARPES, especially strong near the FS edges in $\alpha$ along the surface normal direction.
There is also a Eu $4d \rightarrow 4f$ resonance at $h\nu > 135$~eV, which enhances enormously a Eu 4$f$ peak at 1.5~eV which may distort near-\EF\ features.
Anyhow, from these results, we can see that ARPES spectra at $h\nu = 130$~eV provide information on the electronic structure fairly close to the $X$ point, and because of its 3D nature, its electronic structure cannot be accounted for by a surface state.

\begin{figure}[t]
\includegraphics[width=\columnwidth]{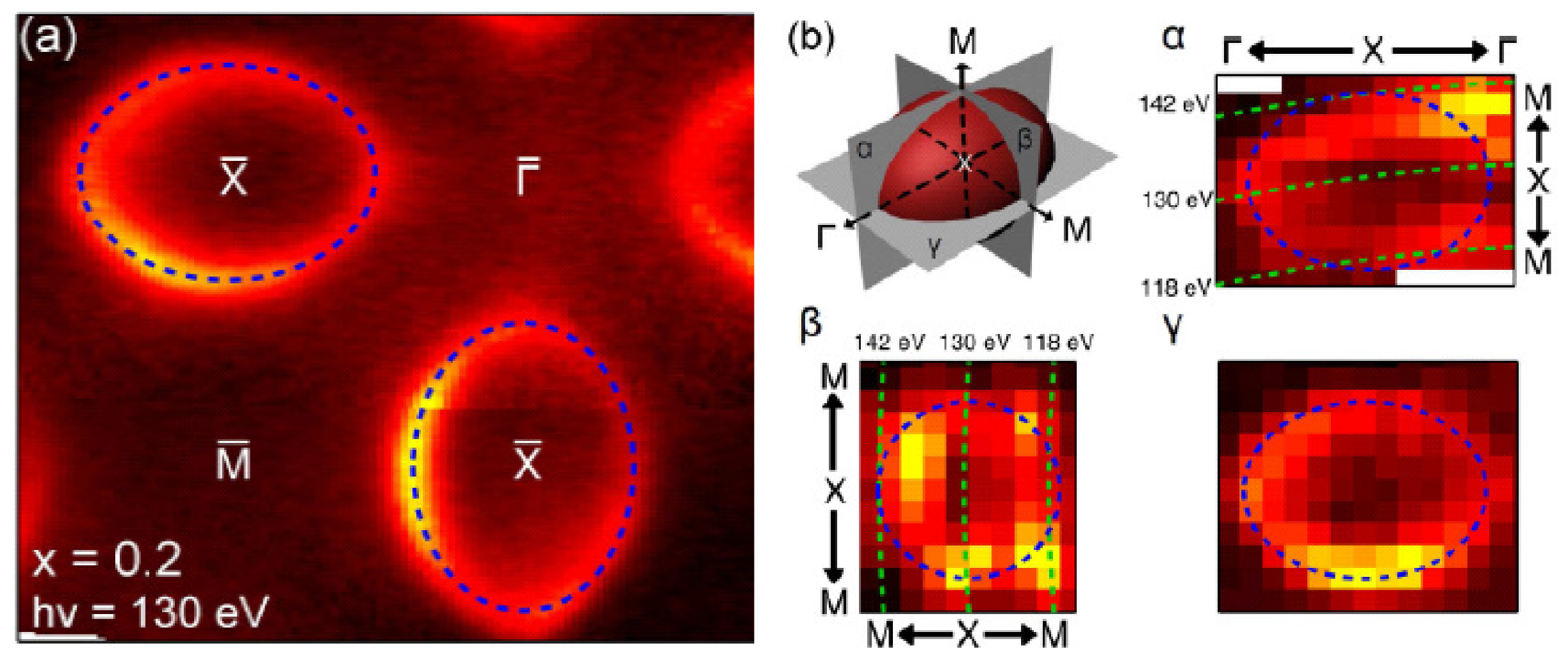}
\caption{
(a) FS map of \ELBE\ taken at $h\nu = 130$~eV and $T = 40$~K.
(b) Cross-sectional FS maps of the same sample taken from $h\nu = 115$ to 146~eV.
ARPES spectra are measured along the green dashed lines at each photon energy, and blue dashed elliptical guidelines for the FSs are of the same size.
}
\end{figure}

Regarding the issue of the bulk band gap between the Eu 5$d$ and the B 2$p$ bands at the $X$ point \cite{Denlinger02}, we investigated ARPES spectra of \ELBE\ around the $X$ point, but it was hard to observe the B 2$p$ band, which may be due to the matrix-element effect in ARPES and strong Eu 4$f$ signals near the Eu 5$d$ band minimum.
Thus, we measured ARPES spectra in a more extended BZ, and Fig.~2 shows the results.
Figure~2(a) is again its FS map, and \Gn, \Xn, and \Mn\ denote the $n$-th nearest high symmetry points with respect to the sample surface normal.
As clearly seen in the map, the intensities around the \Xone\ and \Xtwo\ points are much different from each other.
Figure~2(b) is a constant-energy map at the binding energy ($E_B$) of 0.7~eV, which shows striking difference not only in its intensity but also in shape, i.e. we can see a hollow at \Xone\ but an intense spot at \Xtwo.

To see the origin of these differences, we obtained ARPES spectra along the green dashed line in Figs.~2(a) and 2(b), and present them in Fig.~2(c).
Since Eu$^{2+}$ 4$f$-peak signals are huge around $E_B = 1.5$~eV because Eu 5$d$ and B 2$p$ photoionization cross-sections are about two orders of magnitude smaller than that of Eu 4$f$  \cite{Yeh}, it is hard to identify clear Eu 5$d$- and B 2$p$-band dispersions in a usual ARPES-intensity plot.
Thus, we presented them in a log scale by taking the logarithm of the ARPES intensity of a momentum distribution curve (MDC) after dividing by its minimum.
Then, we can not only enhance very low ARPES signals but also effectively remove momentum-independent ones, especially the tails of the Eu 4$f$ peaks.
As clear in the figure, near the \Xone\ point only the Eu 5$d$-band signals are dominant and there seems to be no sign of the B 2$p$ band.
In contrast, near the \Xtwo\ point the B 2$p$-band signals are stronger than the Eu 5$d$ ones.
Thus, we can conclude that ARPES spectra should, at least in our ARPES geometry and at $h\nu = 130$~eV, be taken around the \Xtwo\ point to observe correctly the B 2$p$ band dispersion.

Another interesting point is that the B 2$p$ band overlaps by about 0.6~eV the Eu 5$d$ band around $E_B = 0.7$~eV, which strongly suggests that intrinsic EuB$_6$ should be a compensated semimetal if both bands are rigidly shifted by La doping.
Recent, ARPES measurements of CeB$_6$ also show the similar overlap of the Ce 5$d$ and the B 2$p$ bands at the $X$ point, which is well described by theoretical bulk band structure \cite{Neupane}, thus strengthening our argument. 

\begin{figure}[t]
\includegraphics[width=0.8\columnwidth]{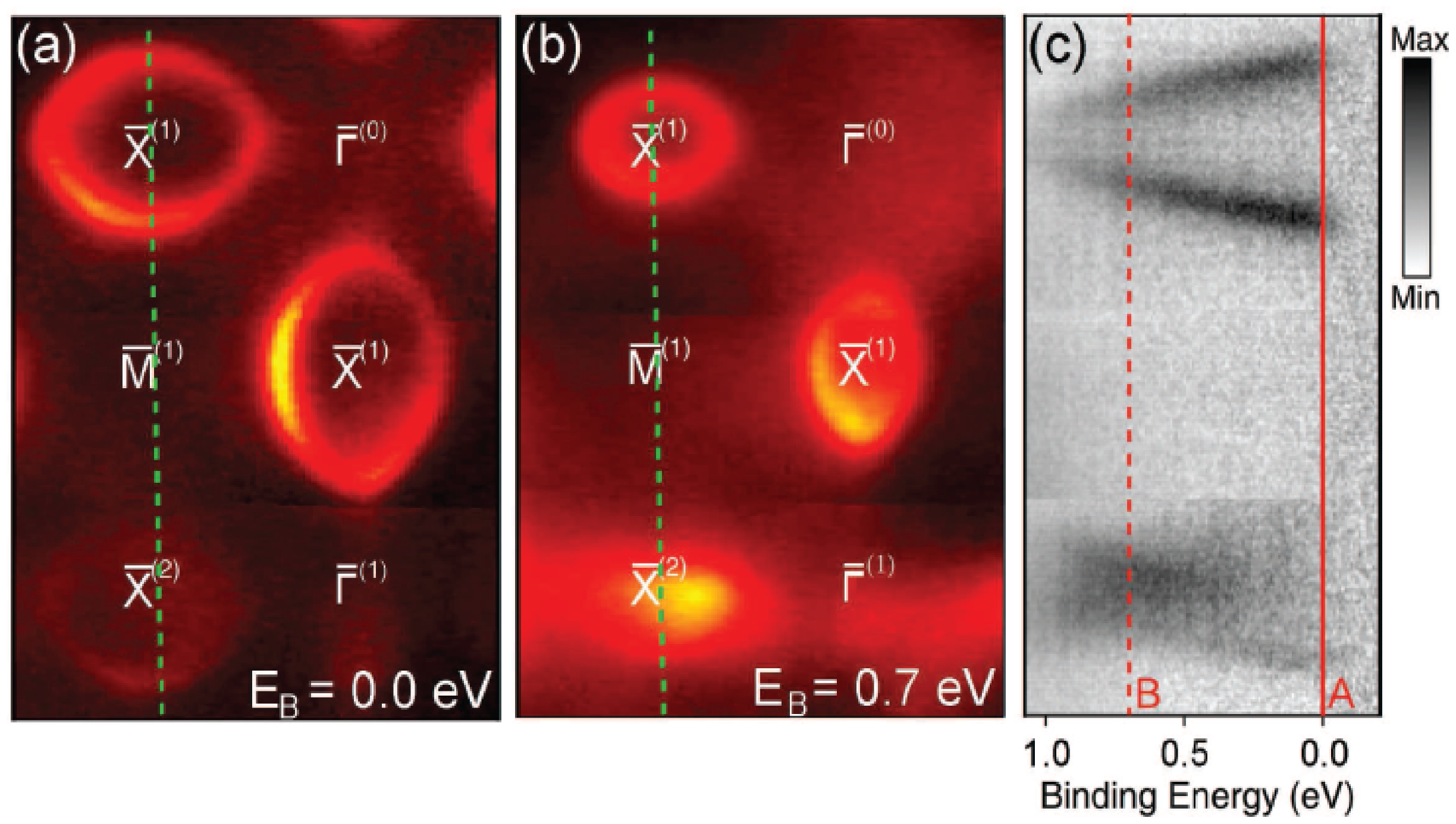}
\caption{
(a) FS map of \ELBE\ in an extended BZ.
\Gn, \Xn, and \Mn\ denote the $n$-th nearest high-symmetry points with respect to the normal direction of the sample surface.
(b) Same as in (a) but a constant-energy-surface map taken at $E_B = 0.7$~eV.
(c) Log-scaled ARPES spectra along the green dashed line in (a) and (b).
Red solid and dashed lines denoted by A and B show where the maps (a) and (b) are obtained, respectively.
}
\end{figure}

Figure~3(a) shows the FS map of EuB$_6$ in the PM phase at $T = 40.7$~K.
As expected, we can observe a weak electron pocket around the \Xone\ point (Eu 5$d$ signals vanish as approaching the \Xone\ point as seen in Fig.~2(c)), the volume of which is nearly a half of that by band calculations for an antiferromagnetic (AFM) phase to mimic the PM one \cite{Shim}.
Around the \Xtwo\ point, there is a more intense spot in contrast to Fig.~2(a), which implies that its character is different from that around the \Xone\ point.
The FS map of EuB$_6$ is quite similar to the map in Fig.~2(b).
Figure~3(b) is a log-scaled ARPES image measured along the green dashed line in Fig.~3(a), which unambiguously shows that the B 2$p$ band crosses \EF.
Thus, together with the result around the \Xone\ point, we can conclude that the intrinsic EuB$_6$ is a compensated semimetal in which the Eu 5$d$ and the B 2$p$ bands overlap each other at the $X$ point. 

\begin{figure}[t]
\includegraphics[width=0.6\columnwidth]{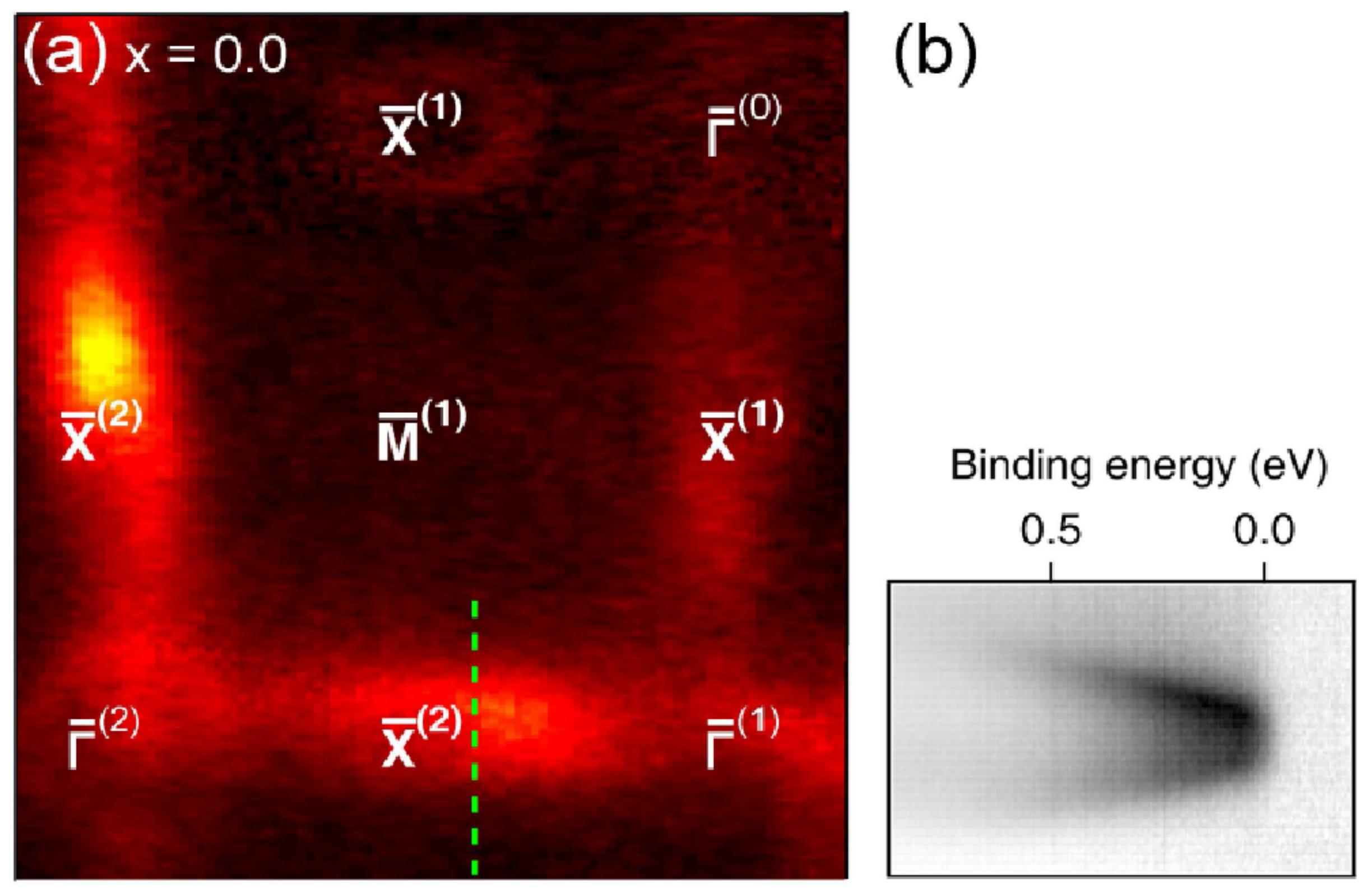}
\caption{
(a) Same as in Fig.~2(a) but the sample is EuB$_6$. 
Interestingly, the FS map is quite similar to the constant-energy map in Fig.~2(b) of \ELBE.
(b) Log-scaled ARPES spectra along the green dashed line in (a).
}
\end{figure}

To examine the electronic-structure change of EuB$_6$ across the FM transition, we obtained temperature-dependent ARPES spectra along the green dashed line in Fig.~3(a), and present them in Fig.~4 upper panels.
Yellow dashed lines are drawn by a parabola with an effective mass of 0.19 $m_e$, which match the peak positions in energy distribution curves and MDCs.
The effective mass is also comparable to the value of 0.25 $m_e$ obtained by dHvA measurements obtained in the minor axis of the elliptical hole pocket \cite{Goodrich}.
To be sure of spectral behavior described below, we also present MDCs taken at $E_B = 0.0$ and 0.3~eV in lower panels.

As the temperature decreases near to \TC\ ($\sim15.5$~K), we can observe that the maximum of the B 2$p$ band gradually shifts downward leaving some diffuse structure near \EF\ denoted by red triangles.
This behavior is more clearly seen in the lower panel in which peaks denoted by red triangles become stronger with decreasing temperature.
Peak positions for the down-shifted band are denoted by blue triangles.
At $T = 15.7$~K and $E_B = 0.3$~eV, we can see a shoulder denoted by a red triangle that seems to form a band below \TC.
Well below \TC, this structure forms a well-defined band to show similar
dispersion to that of the down-shifted band.

The separation between two bands is about 0.35~eV, the origin of which should be the exchange interaction between the valence band and the local 4$f$ moments that have an FM order, as observed in the ARPES spectra of Gd metal \cite{Kim92}.
Since the B 2$p$ band has an AFM Kondo coupling to the Eu 4$f$ states \cite{Kunes}, we can assign the upper/lower band as the majority/minority-spin band denoted by red/blue arrows in the lower panels, and their separation is comparable to theoretical estimations \cite{Kunes,Shim}.

Returning to the diffuse structure near \EF\ above \TC, now we can see that it is a vestige of the exchange-split majority-spin band and eventually merges with the minority one at $T = 40.7$~K.
Since the separation of the exchange-split bands should be proportional to sample magnetization \cite{Kreissl,Nolting85}, if the PM phase is magnetically homogeneous or if the density of itinerant electrons is spatially uniform even when there are FM islands in the PM phase, there should be no difference between spin-up and -down electrons by symmetry, i.e. no exchange splitting.
However, if there is a small FM island, B 2$p$ majority/minority-spin electrons feel local repulsive/attractive potential by AFM exchange interactions and may form a (virtual) anti-bound/bound state, i.e. a MP, which mainly resides near the B 2$p$-band maximum/minimum in the BZ, if the potential is not so strong 

\onecolumngrid
\begin{center}
\begin{figure}[t]
\includegraphics[width=\textwidth]{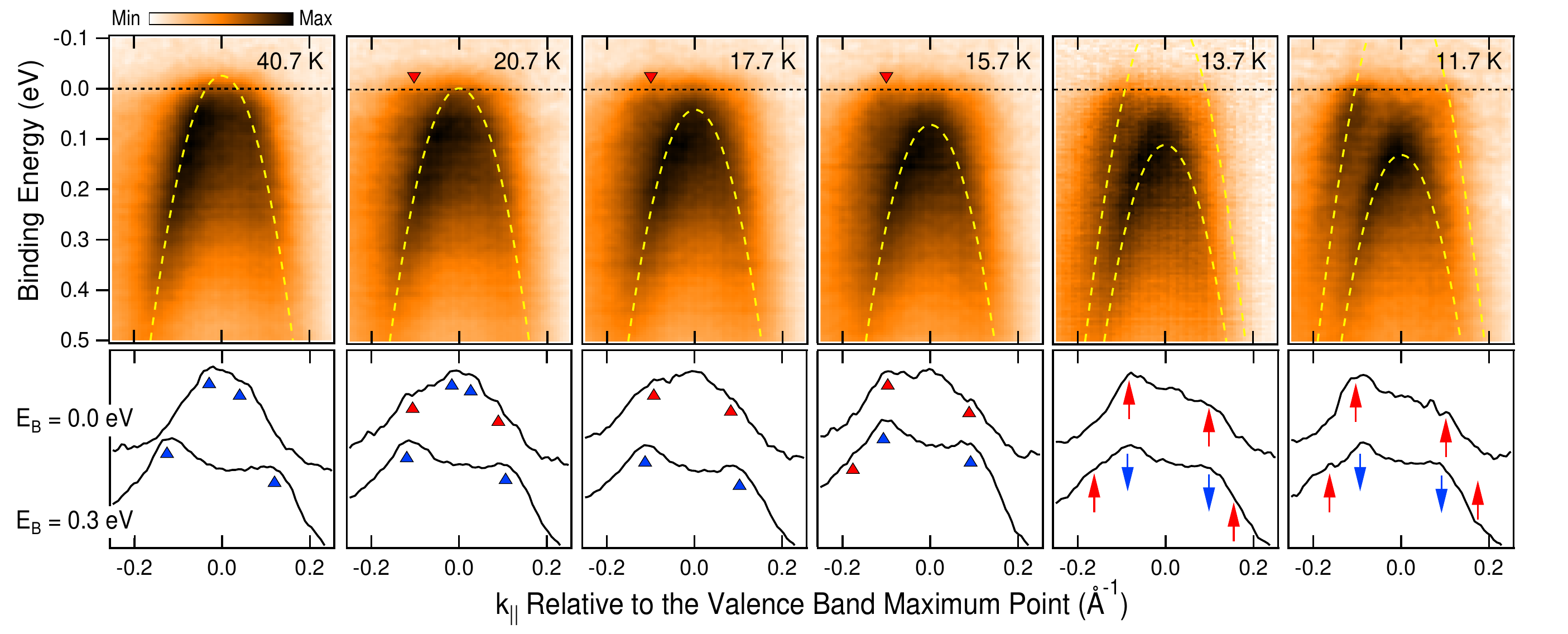}
\caption{
Upper panels: Log-scaled ARPES spectra of EuB$_6$ along the green dashed line in Fig.~1(a) by varying the temperature from 40.7~K to 11.7~K.
Red triangles indicate spectral features which may be related to MPs, and yellow dashed parabolas are guidelines with an effective mass of 0.19 $m_e$.
Lower panels: MDCs taken at $E_B = 0.0$ and 0.3~eV to show spectral features in the upper panels more clearly. Red triangles are peak positions for MPs, blue triangles for a merged band, and red/blue arrows for an exchange-split majority/minority band.
}
\end{figure}
\end{center}
\twocolumngrid
\noindent
as in our case where the exchange interaction is an order of magnitude smaller than the B 2$p$-band width.
Then, in ARPES spectra of the majority-spin band near the $X$ point, we can expect a diffuse anti-bound state and a slightly down-shifted band, the separation between which increases as does the size of the FM island, while in other $k$ region the separation between the majority- and minority-spin bands is negligible.
In the PM phase, this behavior should take place both in the majority- and minority-spin bands and ARPES features above \TC\ in Fig.~4 can be qualitatively explained by this scenario.\vspace*{15mm}

To describe quantitatively the temperature-dependent spectral behavior, we calculated ARPES spectra using a 2D Kondo lattice model \cite{Yu06}
\begin{eqnarray}
H &=&-t \sum_{<i,j>\sigma} c^\dagger_{i\sigma} c^{  }_{j\sigma}
              -J \sum_{i} \vec{\sigma}_i \cdot \vec{S}_i,
\end{eqnarray}
where $c^\dagger_{i\sigma}$ creates a conduction electron with spin $\sigma$ at site $i$, and $\vec{\sigma}_i$ and $\vec{S}_i$ are the spins of a conduction electron and a local magnetic moment, respectively.
$t$ is a nearest-neighbor conduction-electron hopping parameter and $J$ an exchange coupling between the conduction-electron spin and the local moment.
$<i,j>$ denotes a nearest-neighbor pair in the 2D square lattice. 
Though EuB$_6$ is a 3D system, the dimensionality may not be so important because the size of a MP is less than 10 unit cells as shown below.
For the local moments, Ising spins ($\vec{S}_i = \pm 1/2 \hat{z}$) are assumed to diagonalize exactly the model Hamiltonian for a large-size lattice.
In order to mimic the B 2$p$ band dispersion at \EF\ and the exchange splitting,  $t = 1$~eV and $J = 1.2$~eV were assumed, which belongs to the weak-coupling regime in contrast to \cite{Yu05,Yu06}.
To have a similar FS volume, calculations were carried out for a 1\% hole-doped system.

The model Hamiltonian was solved by the Monte Carlo method using the Metropolis algorithm for a $N \times N$ 2D square lattice ($N = 40$) with an open boundary condition for a better momentum resolution than with a periodic one.
Figure 5(a) shows local-moment distributions on the lattice at several temperatures, which was averaged over $N^2$ configurations after equilibrium was reached, 
and Fig.~5(b) their average values $S_{av}$ according to the temperature.

We can see that the FM transition occurs around $T/t = 0.0075$, which is not sharply defined because of finite-size effects.
Interestingly, even when $T$ is two times larger than \TC, the averaged local moment at each lattice site is far from the zero value and there are spin-up or -down FM islands of which size are about five unit cells.
Figure 5(c) shows hole-density distributions which correspond to Fig.~5(a).
For $T >$ \TC, the holes are both on spin-up or -down FM islands.
Below \TC, however, they are mainly on spin-down FM islands, which is understood by the AFM Kondo coupling between the conduction-electron spin and the local moment, which results in partially unoccupied spin-up and fully occupied spin-down bands.
For comparison, a hole-density distribution at $T = 0$ is shown in Fig.~5(d), in which the holes are not 

\onecolumngrid
\begin{center}
\begin{figure}[t]
\includegraphics[width=\textwidth]{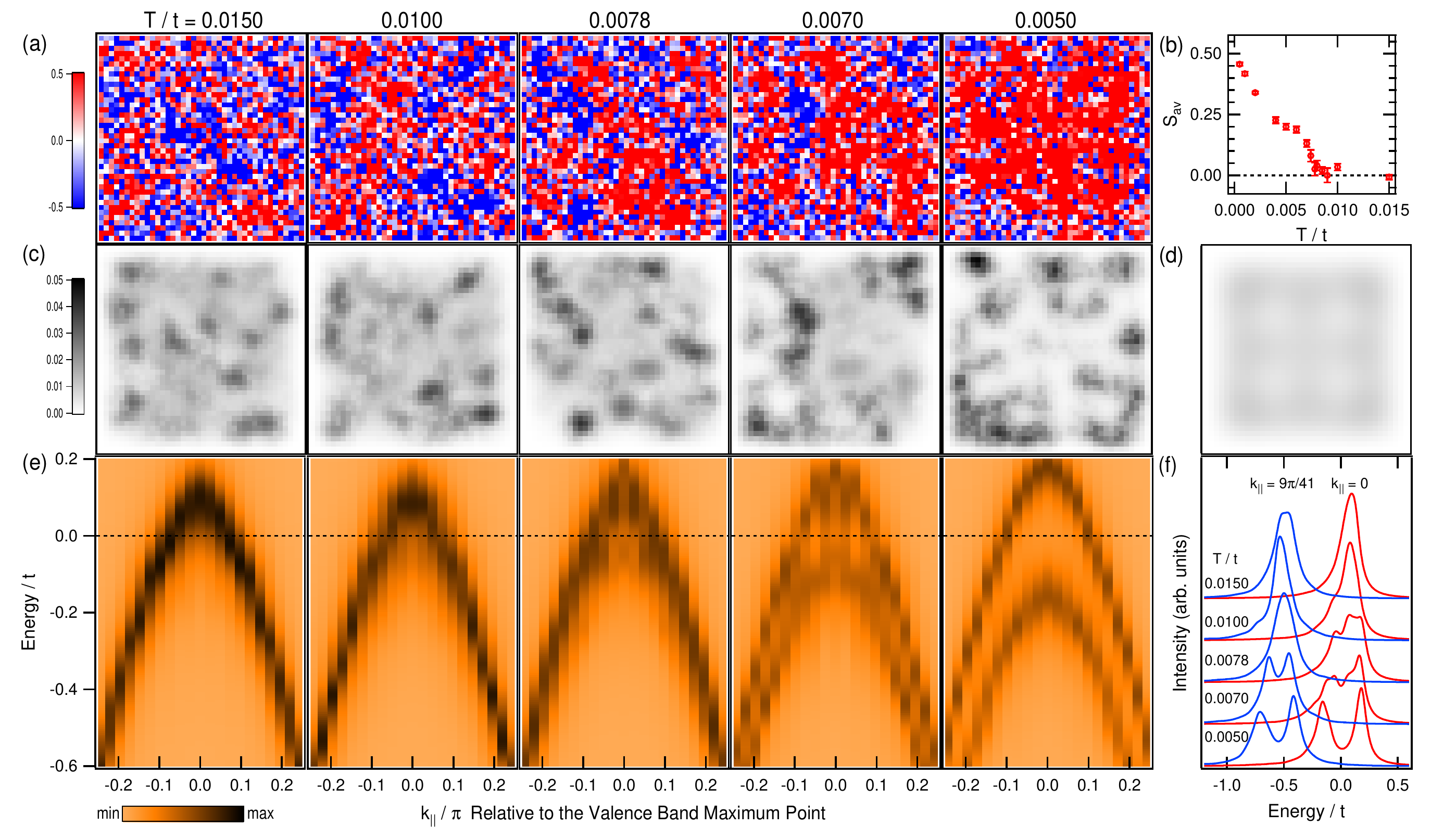}
\caption{
Results of Monte Carlo simulations for a 1\%-hole-doped $40\times 40$ 2D square Kondo lattice model with an open boundary condition:
(a) Local-moment distributions on the lattice according to the temperature averaged over 1,600 configurations after equilibrium was reached.
(b) Average of local moments according to the temperature, which shows \TC~$ \sim 0.0075 t$. 
(c) Hole-density distributions which correspond to Fig.~5(a).
(d) Ground-state conduction-hole distribution at $T = 0$.
(e) Snapshots of ARPES images near the conduction-band maximum which correspond to Fig.~5(a).
ARPES spectra are broadened by a Lorentzian of $0.1 t$ full width at half maximum.
(f) Energy distribution curves at $k_\parallel = 0$ (red) and $9\pi/41$ (blue) relative to the conduction-band-maximum momentum.
}
\end{figure}
\end{center}
\twocolumngrid
\noindent
uniformly distributed because of the open boundary condition but do not make a highly concentrated island as in Fig.~5(c).
All these facts strongly suggest that MPs are robust above \TC\ in this model system.

Figure 5(e) shows snapshots of ARPES images near the conduction-band 
maximum which corresponds to Fig.~5(a) \cite{comment}.
To compare with experimental ones in Fig.~4, they are plotted symmetrically in $k_\parallel$.
At high-enough temperature, we can see a single-band crossing \EF, and well below \TC, exchange-split two bands with near-uniform separation for all momenta.
As the temperature increases, the separation between the two bands becomes much smaller well below \EF\ than near the conduction-band maximum, and nearly merges and leaves a distinctive structure near \EF\ (see third and fourth ARPES images in Fig.~5(e)).
To see this spectral behavior more clearly, we depicted energy distribution curves in Fig.~5(f) at $k_\parallel = 0$, and $9\pi/41$ relative to the conduction-band-maximum momentum. 
All these calculated spectral features are quite similar to the experimental ones in Fig.~4, and consistent with the qualitative argument in the previous section, thus the spectral behavior in ARPES above \TC\ can be explained by a MP formation.

ARPES measurements of EuB$_6$ in an extended BZ show unambiguously that the B 2$p$ band crosses \EF, thus proving that EuB$_6$ is a semimetal. Temperature-dependent ARPES spectra show clear exchange-split bands in the FM phase, and even in the PM phase, there is an exchange-driven diffuse structure near \EF, which provides evidence for the formation of MPs. Such spectral behavior is supported by Monte Carlo simulations of ARPES spectra for a 2D Kondo-lattice model.

\begin{acknowledgments}
We appreciate helpful discussions with B. I. Min, E. J. Choi, J. D. Denlinger and J. W. Allen.
HDK was supported by National Research Foundation of Korea (NRF) grant funded by the Ministry of Science and ICT (No. NRF-2021R1F1A1063881) and BK and BGC by GIST in 2021 and by NRF (No. NRF-2017R1A2B2008538).
\end{acknowledgments}


\begin{thebibliography}{}
\bibitem{comment}
Usual ARPES-measurement time is several orders of magnitude smaller than the time for flipping a local moment in a Kondo lattice model.
\end{thebibliography}
\end{document}